# Near-to-far field transformations for radiative and guided waves

Jianji Yang[1*,#], Jean-Paul Hugonin[2*] and Philippe Lalanne[1*]

1 Laboratoire Photonique Numérique et Nanosciences (LP2N, UMR 5298), Institut d'Optique d'Aquitaine, CNRS - IOGS - Univ. Bordeaux, 33400 Talence, France.
2 Laboratoire Charles Fabry (LCF, UMR 8501), CNRS - IOGS - Univ. Paris-Sud, 2 Avenue Augustin Fresnel, 91127 Palaiseau, France

**Abstract**

Light emitters or scatterers embedded in stratified media may couple energy to both free space modes and guided modes of the stratified structure. For a comprehensive analysis, it is important to evaluate the angular intensity distribution of both the free-space modes and guided modes excited in such systems. In the present work, we propose an original method based on Lorentz-reciprocity theorem to efficiently calculate the free-space and guided radiation diagrams with a high accuracy from the sole knowledge of the near-field around the emitters or scatterers. Compared to conventional near-to-far field transformation techniques, the proposal allows one to easily evaluate the guided-mode radiation diagrams, even if material dissipation is present in the stack, and thus to simultaneously track the coupling of light to all channels (i.e., free-space and guided-ones). We also provide an open-source code that may be used with essentially any Maxwell's equation solver. The numerical tool may help to engineer various devices, such as light-emitting diodes or nanoantennas to achieve directional and efficient radiative spontaneous decays in free space and guided optics.

**Key words:** light emission, light scattering, mode expansion, radiation diagram, stratified structure, Lorentz reciprocity, computational electrodynamics

**Introduction**

When excited either by propagative electromagnetic waves (e.g., plane waves or guided waves) or by local sources such as dipole emitters, individual scatterers or aggregates of scatterers embedded in a stratified medium may couple light into multiple channels including guided modes of the stratified structure and free-space radiative modes of the claddings. The coupling mechanism plays a central role in a large variety of research topics in photonics, for instance for improving the performance of thin-film solar cells using metal nanoparticles [1, 2], enhancing light extraction in light emitting devices using nanoparticles [3-5], achieving directional light scattering [6,7] or emission [8-10] using nanoantennas [11], realizing various optical functionalities (e.g., beam steering [12-13] and shaping [14]) with metasurfaces based on nanoresonator arrays, or engineering the scattering properties of plasmonic nano-objects such as metallic slits/grooves [15-17] and metallic holes [18-20].

In electromagnetic simulations with Maxwell's equation solvers, usually one can easily compute only the near-zone field around the scatterers. However, the far-zone field is often of great interest because it contains all the information on the so-called *free-space radiation diagram*, which describes the angular distribution of the energy flux density radiated into free space, and the *guided-mode radiation diagram*, which describes the angular distribution intensity radiated into guided modes of the stratified medium. Numerical techniques capable of computing far-zone radiation diagrams only using near-zone field distributions are usually referred to as near-to-far field transformation (NFFT) approaches. The latter are mature techniques in microwave engineering [21]. Nearly all numerical implementations of NFFT techniques rely on the field equivalence principle (or Huygens' principle) [21], which allows us to replace the radiated field of an antenna in the near-field zone by equivalent current sources, and then to generate far-zone fields using the free-space Green's function and the equivalent sources. For antennas placed in a stratified substrate, the free-space Green's function must be replaced by Green's functions that account for the strata [22,23].

Many authors have contributed to the elaboration of numerical methods on NFFT approaches, in various fields of physics, maybe starting from initial works on the emission and propagation of radio waves above the earth [24]. However it appears that the literature only contains scattered electromagnetic contributions and that a unified numerical tool that retrieves both free-space and guided-mode radiation diagrams is still lacking, at least in the optical domain [23]. Hereafter, we remove this deficiency and establish a formalism to retrieve the far-zone field from the near-zone field for electromagnetism problems due to the scattering by *local inhomogeneities* in stratified media. We also provide an open-source NFFT toolbox [25], which retrieves the far-field radiation diagrams from the knowledge of the permittivities and permeabilities of the layers (assumed uniform and isotropic) of the stratified structure and of the near-zone field that can be obtained with virtually any Maxwell's equation solver on a closed surface surrounding the local inhomogeneity.

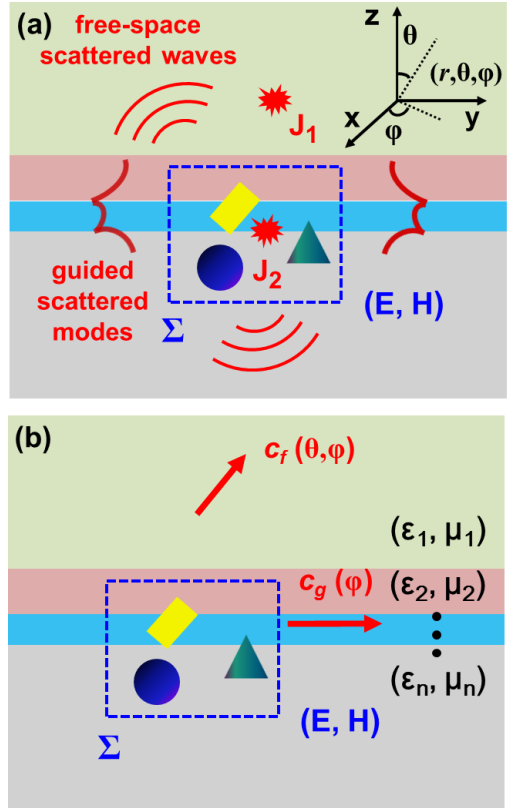

**Figure 1. Radiation diagram calculation due to the scattering of local inhomogeneities in a stratified medium.** (**a**) The scatterers (rectangle, sphere, and triangle) may be excited by either a far-field source $J_1$ or a near-field source $J_2$. They excite both free-space radiative modes and guided modes of the stratified structure. (**b**) Retrieval of the angular distribution of plane-wave amplitudes $c_f(\theta,\varphi)$ and guided-mode amplitudes $c_g(\varphi)$. The blue closed "box" $\Sigma$ fully encloses all the *local inhomogeneity*, so that the geometry outside the box should be the stratified medium. The retrieval of $c_f(\theta,\varphi)$ and $c_g(\varphi)$ requires the knowledge of the field $(\mathbf{E},\mathbf{H})$ on the "box", the relative permittivity and permeability of every layer. In practice, a small-sized box can be used and $(\mathbf{E},\mathbf{H})$ on the box can be obtained with any Maxwell's equation solver. The accompanying open-source code performs rapid computation of the radiation diagrams (amplitude and intensity) of the free-space radiated waves and the guided waves. Inset in (**a**) shows the coordinates.

To be more precise, let us consider a planar stratified medium, in which a few scatterers are embedded (see Fig. 1a). We further consider a parallelepipedic box that fully encloses the scatterers, so that the geometry outside the box is strictly the stratified medium without any inhomogeneity. The blue closed "box" $\Sigma$ of Fig. 1a provides an example. The inhomogeneities might be illuminated by an external source, such as **J**$_1$ in Fig. 1a, which is placed outside the box (if the source is very far then the illumination is a plane wave) or by a localized source, such as **J**$_2$ in Fig. 1a, which is placed inside the box. We assume that with some Maxwell's equation solver, one is capable to calculate the electromagnetic field (**E**,**H**) on the box. (**E**,**H**) can be either:

- for **J**$_2$ = 0, the scattered field induced by the inhomogeneities and obtained as the difference between the total field and the background field (which is generated by the source **J**$_1$ illuminating the stratified medium in the absence of the inhomogeneities);
- for **J**$_1$ = 0, the total field generated by the source **J**$_2$ that illuminates the embedded scatterers (this situation represents a typical extraction problem encountered with light emitting devices);
- any linear combination of the two previous cases.

For all cases, (**E**,**H**) on the box is formed only by outgoing (regarding the "box") plane waves and outgoing guided modes (no incoming fields) and the present NFFT method and open-source toolbox may accurately retrieve the far-field radiation diagrams for guided and free-space modes from the sole knowledge of (**E**,**H**) and the permittivities and permeabilities of the layers (assumed uniform and isotropic) of the stratified structure. The approach is very general as it applies to inhomogeneities with arbitrary shapes, any materials, and any thin film stack, even for lossy dielectric or metallic materials. The only requirement is that all material must be reciprocal, i.e., $\mathbf{\epsilon} = \mathbf{\epsilon}^T$ and $\mathbf{\mu} = \mathbf{\mu}^T$, where the superscript "$T$" denotes the transpose operator.

The manuscript is organized as follows. Firstly, we briefly describe a general NFFT method for calculating the free-space radiation diagrams. Then, as the main contribution of the present work, the formulation of a novel and substantially analytical method to retrieve guided-mode radiation diagrams is proposed. Thirdly, the efficiency and applicability of the numerical tool is exemplified by studying some basic light scattering/emission problems in stratified media, and the validity and accuracy of the open-source code is tested. In the last section, the usage of the open-source tool and the conditions under which the tool may be used safely are summarized.

**A classical approach to retrieve the free-space radiation diagrams**

In this section, we describe a NFFT method to retrieve the far-zone free-space radiation diagrams. This is a classical problem in electromagnetism, which has already been treated by many authors, see for instance the recent works in [22, 23], based on a Green's dyadics approach and the field equivalence principle. A largely equivalent approach based on Lorentz reciprocity theorem is briefly outlined hereafter. This is not the main focus of the present work, however, we think it is still worth being documented concisely for the sake of completeness, since it is implemented in the open-source NFFT tool and is inherently linked to the original contribution, the NFFT of guided-modes, which is presented in the next section.

We consider the radiation diagram into the upper half-space ($z \rightarrow \infty$) of the field distribution $(\mathbf{E}, \mathbf{H})$, assuming $\text{Im}(\varepsilon_1) = 0$ and $\text{Im}(\mu_1) = 0$. The radiation in the lower half-space is treated similarly. Since the radiation diagram essentially describes the angular distribution of the energy flux density, the retrieval consists in performing a plane wave decomposition of $(\mathbf{E}, \mathbf{H})$ and to compute the amplitudes of the radiated plane waves for every direction. Both TE- and TM-polarized waves have to be taken into account for every direction, as implemented in the open-source code; however, because plane-waves with orthogonal polarizations are independently retrieved, in the following we consider purely TE- or TM-polarized waves to reduce cumbersome notations.

It is convenient to first consider the upward-going plane wave (either TE or TM polarized) that propagates away from the scatterers in the directions defined by $(\theta, \varphi)$, see the inset in Fig. 1a. The plane-wave is denoted by $(\hat{\mathbf{E}}_0^+, \hat{\mathbf{H}}_0^+) \exp[i(ux + vy + \chi z)]$, with the superscript '+' of $\hat{\mathbf{E}}_0^+$ and $\hat{\mathbf{H}}_0^+$ labeling upward-going waves (a minus sign is used hereafter for downward-going waves), and $u = k_0 \sqrt{\varepsilon_1 \mu_1} \sin(\theta) \cos(\varphi)$, $v = k_0 \sqrt{\varepsilon_1 \mu_1} \sin(\theta) \sin(\varphi)$, $\chi = k_0 \sqrt{\varepsilon_1 \mu_1} \cos(\theta)$, $k_0$ denoting the wave-number in vacuum. In an horizontal plane at $z = z_1$ in the upper half-space, $(\mathbf{E}, \mathbf{H})$ can be expanded as

$$\mathbf{E}(x, y, z_1) = \iint c_f^+(\theta, \varphi) \hat{\mathbf{E}}_0^+ \exp[i(ux + vy + \chi z_1)] du dv, \qquad (1.1)$$

$$\mathbf{H}(x, y, z_1) = \iint c_f^+(\theta, \varphi) \hat{\mathbf{H}}_0^+ \exp[i(ux + vy + \chi z_1)] du dv, \qquad (1.2)$$

where $c_f^+(\theta, \varphi)$ denotes the plane wave amplitude coefficient. In principle, $c_f^+(\theta, \varphi)$ can be found with inverse Fourier transform. However this requires the knowledge of the radiated field in the entire (infinite) plane $z = z_1$, which is not suitable for numerical calculations.

Indeed, according to Lorentz reciprocity theorem, $c_f^+(\theta, \varphi)$ can be obtained by computing a surface integral on a closed "box" $\Sigma$ surrounding the local inhomogeneities (see Fig. 1b)

$$c_f^+(\theta, \varphi) = \frac{k_0 Z_0 \mu_1}{8\pi^2 \chi \hat{\mathbf{E}}_0^+ \cdot \hat{\mathbf{E}}_0^-} \oiint_\Sigma (\mathbf{E} \times \hat{\mathbf{H}}^- - \hat{\mathbf{E}}^- \times \mathbf{H}) \cdot d\mathbf{S}, \qquad (2)$$

where $Z_0$ denotes the vacuum impedance and $(\hat{\mathbf{E}}^-, \hat{\mathbf{H}}^-)$ represents the field that is created by a downward planewave $(\hat{\mathbf{E}}_0^-, \hat{\mathbf{H}}_0^-)\exp[-i(ux + vy + \chi z)]$ impinging onto the stratified medium (without the inhomogeneities). The flux density at direction $(\theta, \varphi)$ can be found as $\frac{1}{2}\sqrt{\varepsilon_0\varepsilon_1/(\mu_0\mu_1)}|c_f^+(\theta,\varphi)|^2|\hat{\mathbf{E}}_0^+|^2$. In passing, we emphasize that, because the vectorial field $(\hat{\mathbf{E}}^-, \hat{\mathbf{H}}^-)$ induced by a planewave in a stratified structure can be calculated rather easily with 2×2 matrix products for every incidence and polarization [22, 26], the overlap integral of Eq. 2 can be computed in parallel for all directions and the $c_f^+(\theta,\varphi)$ coefficients can be computed very efficiently (see **Technical remarks**).

We do not provide a demonstration of Eq. (2) with the Lorentz reciprocity theorem hereafter; it is technical and essentially echoes the classical NFFT techniques based on the field equivalence principle [21], without explicitly considering the equivalent current sources.

**Semi-analytical approach to retrieve the guided-mode radiation diagrams**

In this section, an original NFFT approach for retrieving guided mode radiation diagrams for planar stratified structure is formulated. We again consider the emission problem sketched in Fig. 1b, in which the guided modes of the planar stratified structure may be launched in all in-plane directions [i.e., $0 < \varphi < 2\pi$].

Before going to details, we would like to briefly address the following question: bearing in mind the NFFT method for free-space waves described in the former section, can we extract the guided-mode radiation diagram by performing, similarly to Eq. (2), an overlap integral on the closed "box" $\Sigma$ between the field $(\mathbf{E}, \mathbf{H})$ and the field of the associated to incident guided mode propagating in the direction $\varphi$, i.e. replacing the incident plane wave $(\hat{\mathbf{E}}^-, \hat{\mathbf{H}}^-)$ by an incident guided mode with an in-plane wavevector parallel to the azimuthal direction $\varphi$? The answer that can be easily derived from the Lorentz reciprocity theorem is yes for non-lossy materials, but the treatment is problematic for lossy modes because one needs to handle exponentially large fields.

To remove this difficulty, we formulate a novel and substantially analytical approach that allows one to evaluate the radiation diagram of guided-modes with a high accuracy, for lossy (dielectric or metallic) stratified structures. The method just requires for the knowledge of the field $(\mathbf{E}, \mathbf{H})$ on the closed box $\Sigma$ and of the guided mode profile. All the derivation steps are provided hereafter, with some technical details being documented in the SI.

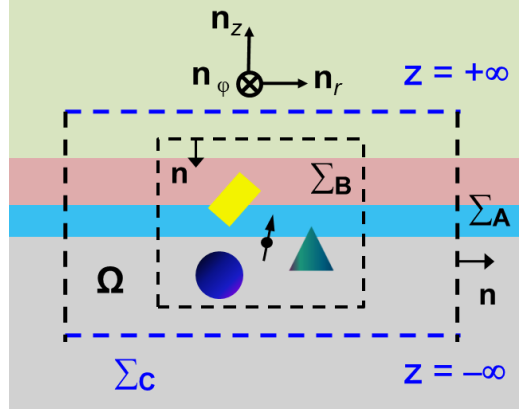

**Figure 2. Domain Ω with its boundaries.** $\Sigma_A$ (vertical black dashed lines) defines a cylindrical surface of radius *r* that extends from $z = -\infty$ to $+\infty$, $\Sigma_B$ is a closed "box" that encloses the inhomogeneity, and $\Sigma_C$ corresponds to the upper and lower horizontal surfaces (blue dashed lines) of the cylinder at $|z| = \infty$. $\Sigma_A$ and $\Sigma_C$ together form the outer boundary of the volume Ω, and $\Sigma_B$ is the inner boundary. The outward surface normal of Ω is denoted by **n**.

**Cylindrical waveguide modes.** We start by expanding the radiated field (**E**,**H**) using waveguide modes of the stratified structure defined in a cylindrical coordinate system $(r,\varphi,z)$, see Fig. 2. When solving source-free Maxwell's equations (eigenmodes problems) of stratified planar waveguides, the eigenmodes can be written as the product of functions with separated variables, *z*, $\varphi$ and *r*, with the *r*-dependent part expressed with Hankel functions (see SI). Hereafter, the $m^{th}$-order TE- or TM-polarized modes with propagation constants $k_m$ will be denoted as $\Phi_{+m,n} = (\hat{\mathbf{E}}^+_{m,n}\exp(in\varphi), \hat{\mathbf{H}}^+_{m,n}\exp(in\varphi))$ for outgoing (to $r \to \infty$) modes propagating and $\Phi_{-m,n} = (\hat{\mathbf{E}}^-_{m,n}\exp(-in\varphi), \hat{\mathbf{H}}^-_{m,n}\exp(-in\varphi))$ for ingoing (to $r \to 0$) modes, with *n* (*n* = 0, ±1…) determining the azimuthal pattern and $(\hat{\mathbf{E}}^\pm_{m,n}, \hat{\mathbf{H}}^\pm_{m,n})$ describing the *r*- and *z*-dependent profiles. Note that due to the separation of variables *r* and $\varphi$, for a given polarization, each order *m* embraces a series of *azimuthal harmonics* with the same propagation constant $k_m$ but with different azimuthal pattern determined by $\exp(\pm in\varphi)$. The explicit expressions of mode profiles (including both the general expression and asymptotic expression at $r \to \infty$) are provided in SI for TE and TM modes.

**Mode expansion.** As in the previous section, we assume that the modes are either TE- or TM-polarized. The total electric field **E** can be decomposed into a set of outgoing guided

cylindrical modes and a continuum of radiation modes

$$\mathbf{E}(r,\varphi,z) = \sum_m \sum_n c_{m,n}^+ \hat{\mathbf{E}}_{m,n}^+(k_m r, z)\exp(in\varphi) + \text{continuum} \ , \tag{3}$$

where $c_{m,n}^+$ denotes the amplitude of the $n^{th}$ *azimuthal harmonic* of the $m^{th}$ outgoing mode for the polarization considered. The second term 'continuum' in Eq. (3) corresponds to a summation over the continuum of radiation modes of the stratified waveguide; we provide some additional discussions in SI about the orthogonality and normalization of radiation modes.

**Asymptotic behavior and radiation diagram**. The electric field of the $m^{th}$ outgoing mode can be expressed as a summation of all the relevant azimuthal harmonics, i.e., $\mathbf{E}_m(r,\varphi,z) = \sum_{n=-\infty}^{+\infty} c_{m,n}^+ \hat{\mathbf{E}}_{m,n}^+(k_m r, z)\exp(in\varphi)$. The asymptotic behavior at $r \to \infty$ can be found with asymptotic forms of Hankel functions and for a TM mode for instance, we have

$$\mathbf{E}_m(r \to \infty, \varphi, z) = f_m(\varphi)\exp(ik_m r)/\sqrt{r}\ [e_L(z)\mathbf{n}_r, 0\mathbf{n}_\varphi, e_T(z)\mathbf{n}_z] \ , \tag{4}$$

where $f_m(\varphi) = \sum_{n=-\infty}^{+\infty} c_{m,n}^+ \exp[-i(2n+1)\pi/4]\sqrt{2/(\pi k_m)}\exp(in\varphi)$ describes the angular distribution of the amplitude of the $m^{th}$ mode in the far zone, $e_L(z)$ and $e_T(z)$ represent the longitudinal and transverse electric field components of the $z$-dependent mode profile (see SI). $|f_m(\varphi)|^2$ characterizes the angular intensity distribution in far-zone, i.e., the *guided-mode radiation diagram*.

If the stack materials are lossless, the total power carried by the $m^{th}$ mode can be found by summing the power of all azimuthal harmonics as $P_m = \int_0^{2\pi}|f_m(\varphi)|^2 d\varphi = 4/|k_m|\sum_{n=-\infty}^{+\infty}|c_{m,n}^+|^2$, provided that the modes are normalized in such a way as $N_{m,n} = (-1)^n 16/k_m$, with $N_{m,n}$ denoting the normalization coefficient of $\Phi_{\pm m,n}$ defined as (see SI)

$$N_{m,n} = \int_0^{2\pi} d\varphi \int_{-\infty}^{+\infty}\left(\hat{\mathbf{E}}_{m,n}^+ \times \hat{\mathbf{H}}_{m,n}^- - \hat{\mathbf{E}}_{m,n}^- \times \hat{\mathbf{H}}_{m,n}^+\right)\bullet\mathbf{n}_r r dz \ . \tag{5}$$

The integral in Eq. (5) runs over a vertical cylindrical surface of radius $r$ with an infinite transverse cross-section from $z = -\infty$ to $z = +\infty$ (e.g., surface $\Sigma_A$ in Fig. 2), and noticeably it involves both the ingoing and the outgoing modes. It is independent of $r$ even in the presence of absorption. Similar normalization approach for guided modes can be found in [27].

**Calculation of the mode amplitude.** Here we will demonstrate that the amplitude coefficient $c_{m,n}^+$ of the mode $\Phi_{+m,n}$ in Eq. (3) can be found with a simple overlap integral [see Eq. (11)] involving $(\mathbf{E},\mathbf{H})$ and the ingoing mode $\Phi_{-m,n}$ on the closed "box" $\Sigma_B$ defined in Fig. 2.

For that, we define the inner product between two modes $\Phi_{\sigma m,n} = \left(\hat{\mathbf{E}}_{m,n}^{\sigma}\exp(\sigma i n\varphi), \hat{\mathbf{H}}_{m,n}^{\sigma}\exp(\sigma i n\varphi)\right)$ and $\Phi_{\xi p,q} = \left(\hat{\mathbf{E}}_{p,q}^{\xi}\exp(\xi i q\varphi), \hat{\mathbf{H}}_{p,q}^{\xi}\exp(\xi i q\varphi)\right)$ (here we use $\sigma$ and $\xi$ to label the propagation direction, i.e., $\sigma = \pm 1$ and $\xi = \pm 1$) as

$$\Phi_{\sigma m,n} \otimes \Phi_{\xi p,q} = \int_0^{2\pi} \exp[i\varphi(\sigma n + \xi q)]d\varphi \int_{-\infty}^{+\infty} \left(\hat{\mathbf{E}}_{m,n}^{\sigma} \times \hat{\mathbf{H}}_{p,q}^{\xi} - \hat{\mathbf{E}}_{p,q}^{\xi} \times \hat{\mathbf{H}}_{m,n}^{\sigma}\right) \bullet \mathbf{n}_r r\, dz, \tag{6}$$

where $\otimes$ denotes the inner product operator. Equation (6) is also defined for a vertical infinite cylindrical surface, as for Eq. (5). Based on Lorentz reciprocity theorem, we establish the unconjugated form of mode orthogonality relation (see SI)

$$\Phi_{\sigma m,n} \otimes \Phi_{\xi p,q} = \delta_{\sigma m,-\xi p}\delta_{n,q}N_{m,n},\ (\sigma = \pm 1, \xi = \pm 1). \tag{7}$$

From Eq. (3) by using the orthogonality relation of Eq. (7), we find

$$c_{m,n}^+ = (\mathbf{E},\mathbf{H}) \otimes \Phi_{-m,n} = \frac{1}{N_{m,n}} \int_0^{2\pi} \exp(-in\varphi)d\varphi \int_{-\infty}^{+\infty} \left(\mathbf{E} \times \hat{\mathbf{H}}_{m,n}^- - \hat{\mathbf{E}}_{m,n}^- \times \mathbf{H}\right) \bullet \mathbf{n}_r r\,dz. \tag{8}$$

As the orthogonality and normalization relations are both *r*-independent (see SI), Eq. (8) is *r*-independent as long as the cylindrical surface, such as $\Sigma_A$ shown in Fig. 2, fully surrounds the inhomogeneity.

Nevertheless, Eq. (8) is not suitable for numerical implementation, since it requires an integral over the entire transverse direction, $-\infty < z < +\infty$. Thus, we consider a closed *source-free* volume $\Omega$ with its boundary $\Sigma$ formed by three surfaces, $\Sigma_A$, $\Sigma_B$, and $\Sigma_C$, see Fig. 2. Applying $(\mathbf{E},\mathbf{H})$ and $\Phi_{-m,n}$ to the Lorentz reciprocity theorem (see SI or Ref. [27]) for the volume $\Omega$, a simple but vital relation can be found as $\oint_\Sigma \exp(-in\varphi)\left(\mathbf{E} \times \hat{\mathbf{H}}_{m,n}^- - \hat{\mathbf{E}}_{m,n}^- \times \mathbf{H}\right) \bullet \mathbf{n}\,dS = 0$, owing to the fact that $(\mathbf{E},\mathbf{H})$ and $\Phi_{-m,n}$ share the same permittivity and permeability distribution in $\Omega$. Furthermore, as we consider the $m^{th}$ mode to be a bound mode with a field that exponentially vanishes at $z = \pm\infty$,

$$\oint_{\Sigma_C} \exp(-in\varphi)\left(\mathbf{E} \times \hat{\mathbf{H}}_{m,n}^- - \hat{\mathbf{E}}_{m,n}^- \times \mathbf{H}\right) \bullet \mathbf{n}\,dS = 0 \tag{9}$$

and we get

$$\int_{\Sigma_A} \exp(-in\varphi)\left(\mathbf{E} \times \hat{\mathbf{H}}_{m,n}^- - \hat{\mathbf{E}}_{m,n}^- \times \mathbf{H}\right) \bullet \mathbf{n}\,dS = -\oint_{\Sigma_B} \exp(-in\varphi)\left(\mathbf{E} \times \hat{\mathbf{H}}_{m,n}^- - \hat{\mathbf{E}}_{m,n}^- \times \mathbf{H}\right) \bullet \mathbf{n}\,dS. \tag{10}$$

The left-hand side of Eq. (10) is the same as the integral in Eq. (8), and therefore finally Eq. (8) can be replaced by

$$c_{m,n}^+ = \frac{1}{N_{m,n}} \oint_{\Sigma_B} \exp(-in\varphi)\left(\mathbf{E} \times \hat{\mathbf{H}}_{m,n}^- - \hat{\mathbf{E}}_{m,n}^- \times \mathbf{H}\right) \bullet (-\mathbf{n}) dS. \qquad (11)$$

Equation (11) constitutes the main result of the present analysis and it can be computed easily on a surface with a finite size (on contrary to Eq. (8)). It implies that the amplitude of an outgoing mode can be formally computed by sending its associated ingoing mode into the stratified medium, as in **Section 2**. We emphasize that the closed "box" $\Sigma_B$ that fully encloses the local inhomogeneity has finite dimensions. Its surface normal **n** is directed inward, see Fig. 2.

We remark that, because the $r$-dependent profiles are depicted by Hankel functions that diverges at $r = 0$, both $\Phi_{+m,n}$ and $\Phi_{-m,n}$ diverge at $r = 0$, however their sum $\Phi_{+m,n} + \Phi_{-m,n}$ remains finite. According to Eq. (7), one can easily find that in Eqs. (8) and (11) $\Phi_{-m,n}$ can be replaced by $\Phi_{+m,n} + \Phi_{-m,n}$, i.e., $c_{m,n}^+ = (\mathbf{E},\mathbf{H}) \otimes (\Phi_{+m,n} + \Phi_{-m,n})$. We emphasize that this replacement, which is strictly rigorous from a mathematical standpoint, is crucial for numerical calculations (additional details can be found in SI).

**Numerical Tests**

In this section, four examples are analyzed to show the capability of the open-source numerical toolbox based on the proposed retrieval approach. Additionally, the validity and accuracy of the numerical tool is tested in examples 1 and 2, for which the radiation diagrams retrieved by the numerical tool from the near-field obtained with COMSOL multiphysics are compared to diagrams obtained analytically. Examples 3 and 4 present two emblematic examples in the area of nanoantennas. Also, implementations of the examples 1 and 2 are provided in Ref. [25].

**Example 1: Scattering by a silicon nanosphere in free space**

We consider a silicon ($n$=3.5) nanosphere of radius $R$ = 285 nm placed in air and illuminated by an $x$-polarized plane wave at a wavelength of 1.5 µm, see Fig. 3a. The radiation diagram of the scattered field is calculated with COMSOL on a rectangular "box" that surrounds the nanosphere and with the present numerical toolbox. The results are displayed with the red curves in Figs. 3b and 3c, and are compared with radiation diagrams obtained with Mie theory [28] that are shown with black circles. The excellent agreement between the fully numerical approach and the analytical solution corroborates the accuracy of the numerical tool for retrieving the free-space radiation diagram.

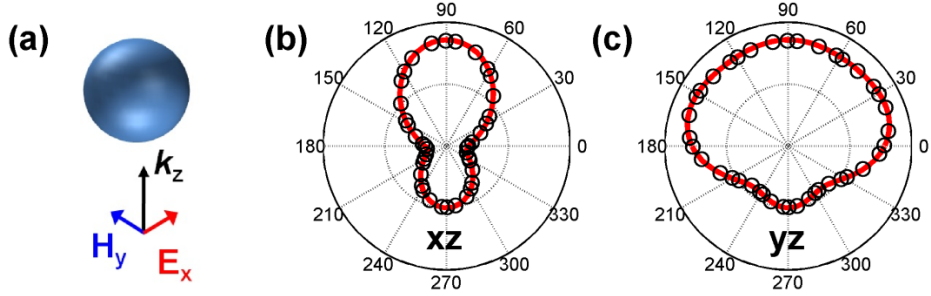

**Figure 3. Planewave scattering by a silicon ($n$ = 3.5) nanosphere (radius $R$ = 285 nm) in air at a wavelength of 1.5 µm**. (**a**) The silicon sphere is illuminated by an $x$-polarized plane wave propagating along +$z$ direction. (**b**) and (**c**) Radiation diagram of the scattered field in $x$-$y$ plane and $y$-$z$ plane. Results obtained with the numerical tool and Mie theory are plotted with red curves and black circles.

**Example 2: emission of two dipole sources in a dielectric slab waveguide**.
In example 2, we consider a more intricate problem, in which two electric dipole sources are placed in a high-index dielectric slab waveguide (see Fig. 4a) deposited on a substrate with a low refractive index. The polarizations of the two electric dipole sources are $\mathbf{J}_1 = \left(\sqrt{2}/2,\ 0,\ \sqrt{2}/2\right)$ and $\mathbf{J}_2 = (0,\ 1,\ 0)$, respectively. The dipoles are separated by 150 nm in the $x$ direction. At the radiation wavelength of 1 µm, the waveguide supports only two modes, $TE_0$ ($n_{eff}$ = 1.24) and $TM_0$ ($n_{eff}$ = 1.20), with all higher order modes being cut-off. The near-field distribution on a rectangular "box" that encloses the two sources is obtained with COMSOL and is used with the numerical toolbox to compute the free-space and guided-mode radiation diagrams. The results are shown in Figs. 4b and 4c-4d with red curves. As the refractive index of the substrate is larger than the one of the superstrate (air), light is dominantly radiated into the substrate. We additionally see that the radiation diagram of TE and TM modes are very different. We have also calculated the radiation diagrams with an analytical expression for the field radiated by a dipole source in a stratified medium using Green's dyadics approaches [23,29,30]. The results shown with black circles are in quantitative agreement with those obtained with the numerical toolbox. Similar agreement has been obtained for the free space radiation diagram (not shown). In addition, we have checked the energy conservation with a relative error < 1%, by comparing the total power radiated by the source doublet computed either directly by the total Poynting flux on a surface surrounding the doublet or by indirectly from the radiation diagrams, summing over all free-space and guided modes.

The excellent agreement between the analytical calculations (black circles) and the fully numerical results (red curves) confirms the effectiveness of the present tool for treating guided modes of the stratified substrate.

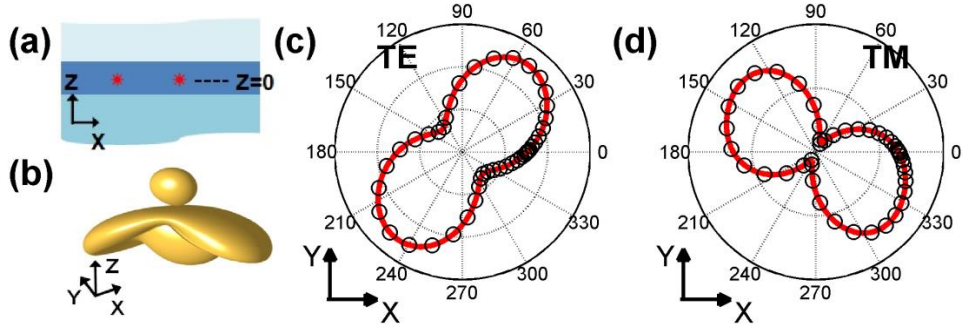

**Figure 4. Emission of two dipole sources inside a dielectric slab waveguide at a wavelength of 1 μm**. (**a**) Sketch of the slab waveguide. The core layer (thickness 200 nm and refractive index 1.5) is surrounded by two semi-infinite media with indices 1 (above) and 1.2 (below). The dipole coordinates are (−100 nm, 0, 0) and (50 nm, 0, 0). (**b**) 3D free-space radiation diagram. The upper spherical (lower three-armed) part corresponds to radiation into the superstrate (substrate). (**c**) and (**d**) Radiation diagrams of $TE_0$ and $TM_0$ guided modes. Results obtained with the numerical toolbox and (semi-)analytical calculation are shown with red curves and black circles, respectively.

**Example 3: Scattering at a metallic hole**
The launching of surface plasmon polaritons (SPPs) by metallic holes or slits has been widely studied in plasmonics [15,16,18-20]. In the third example, we consider the light scattering by a finite-depth rectangular metallic air hole etched in a semi-infinite metal substrate ($z < 0$) and illuminated by a normally incident plane wave, see the inset in Fig. 5a. The hole creates a back scattered field in free space and launches SPPs at the air-metal interface. With the present tool, we quantify the amount of backscattered waves and of launched SPPs. The size of the hole is fixed as $a_x$ = 100 nm and $a_y$ = 400nm. We define the term *generation efficiency* as the ratio of the scattered power (of either free-space waves or SPPs) to the power incident over the hole area ($a_x \times a_y$). The efficiencies are displayed in Fig. 5a, as a function of the hole depth.

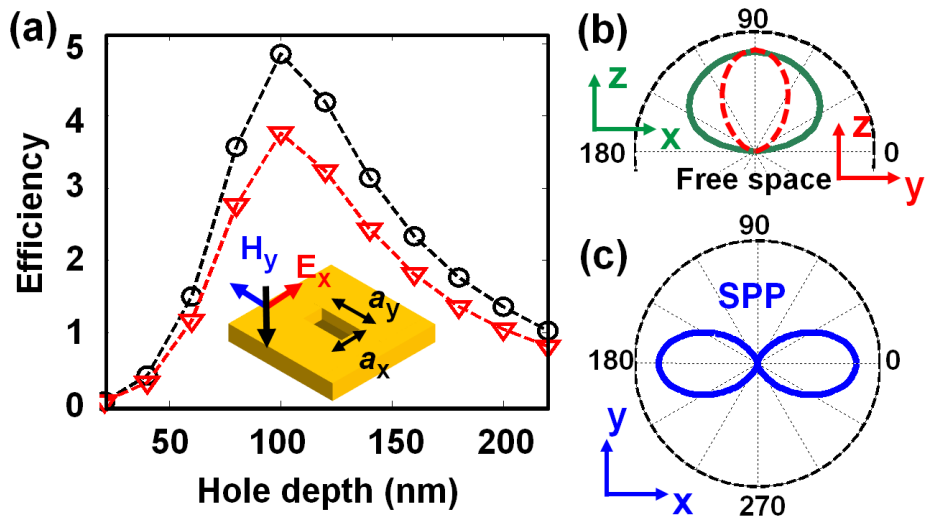

**Figure 5. Scattering by a rectangular metallic hole etched in metal substrate.** (**a**) The hole is illuminated by an *x*-polarized plane wave at $\lambda_0$ = 700 nm impinging at normal incidence, as shown in the inset. Generation efficiencies of free-space scattered waves and SPPs as a function of hole depth are plotted as black circles and red triangles, respectively. (**b**) Free-space radiation diagram of the scattered wave in the *y-z* plane (red dashed) and the *x-z* plane (green solid). (**c**) SPP radiation diagram in the *x-y* plane. The metal permittivity (gold) taken for the simulation is $\varepsilon_m$ = −15.83 + 1.28*i* at $\lambda_0$ = 700 nm.

For the present example, the scattered electromagnetic field on the "box" that encloses the hole is obtained using an aperiodic Fourier modal method [31,32]. Quantitative justification of the retrieved result by comparison with already known results is not made in this case, due to the lack of available solutions for the same problem in the literature. However, the response of sub-wavelength metallic holes can be approximately described as that of an effective magnetic dipole **m**$_y$ [33], and this approximation shows good agreement with the free-space and SPP radiation diagrams shown in Figs. 5b and 5c. The SPPs are dominantly launched toward the *x*-direction, parallel to the incident polarization.

**Example 4: dipole emission inside a metal patch nanoantenna**
Figures 6a-6c depict a metal nanoantenna, composed of a silver nanocube placed on a polymer film coated on a gold substrate and doped with a quantum emitter. Via the nanoantenna, the quantum emitter radiates into free-space and launches SPPs on the dielectric-gold interface, with decay rates denoted by $\gamma_{rad}$ and $\gamma_{SPP}$, respectively. We further denote by $\gamma_{tot}$ the total decay rate. Nanocube antennas have been recently

fabricated and tested, and extremely high Purcell factors and large radiative efficiencies have been reported [10].

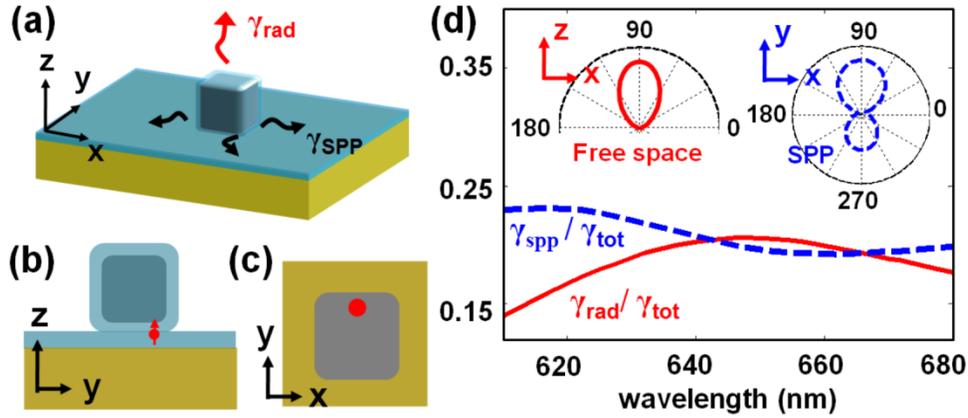

**Figure 6. Dipole emission inside a metal nanoantenna**. (**a**) 3D sketch of the nanoantenna. A silver nanocube (65 × 65 × 65 nm) with a 3-nm polymer coating is placed above a gold substrate covered by a 5-nm polymer (8-nm spacer in total). (**b**) and (**c**) Side and top views of the nanoantenna. A *z*-polarized electric dipole source shown with red dot lies midway between the silver cube and gold substrate, and is located near the edge of the cube. (**d**) Radiative $(\gamma_{rad}/\gamma_{tot})$ and plasmonic $(\gamma_{SPP}/\gamma_{tot})$ quantum-yield spectra. The top-left inset (red solid) shows the free-space radiation diagram in the *x-z* plane at $\lambda_0 = 650$ nm. The top-right inset (blue dashed) shows the SPP radiation diagram in the horizontal *x-y* plane at the same wavelength. In simulation, the refractive index of the polymer is 1.4; for the relative permittivities of silver and gold, a Drude model $\varepsilon_{Ag} = 1 - \omega_{Ag}^2/(\omega^2 - i\omega\Gamma_{Ag})$ and a Drude-Lorentz model $\varepsilon_{Au} = 8.1 - \omega_{Au}^2/(\omega^2 - i\omega\Gamma_{Au}) - 0.7\omega_L^2/(\omega^2 - \omega_L^2 - 2i\omega\Gamma_L)$ are adopted, with $\omega_{Ag} = 1.3\times10^{16}$ s$^{-1}$, $\Gamma_{Ag} = 1.6\times10^{14}$ s$^{-1}$, $\omega_{Au} = 1.3\times10^{16}$ s$^{-1}$, $\Gamma_{Au} = 4.13\times10^{13}$ s$^{-1}$, $\omega_L = 4.1\times10^{15}$ s$^{-1}$, and $\Gamma_L = 4.7\times10^{14}$ s$^{-1}$,.

With COMSOL, we calculate the field radiated by the quantum emitter (modeled as a vertical electric dipole) on a box surrounding the nanocube and infer the radiation diagrams into the SPP and free-space modes. We also calculate the total power radiated by integrating the Poynting flux over a small box surrounding the dipole in the polymer layer. The SPP and free space emission efficiencies are shown in Fig. 6d. The computation is made over a spectral range covering the nanoantenna resonance at $\lambda_0 = 650$ nm. The radiation diagrams at $\lambda_0$ are shown in the insets. Note that at resonance, $\gamma_{rad}/\gamma_{tot}$ is about 20%. A larger value (~50%) was predicted in [10], but we note that the authors in [10] are not equipped with the present numerical tool and that they infer $\gamma_{rad}$ from the difference between the $\gamma_{tot}$ and the non-radiative decay rate $\gamma_{nrad}$ computed as the Joule loss inside the finite computational domain. Such an approach overestimates $\gamma_{rad}$, especially if SPPs that carry energy far-away from the nanoantenna are efficiently launched like in the present case. We believe that this highlights the importance of the present tool. Also, we remark that as there is a fairly large portion of radiation that is directed into SPPs on the dielectric-metal interface, this type of device may be considered as a good candidate for plasmon source [6,34,35]. A detailed study of the decay of quantum emitters in the different decay channels ($\gamma_{tot}$, $\gamma_{rad}$ and $\gamma_{SPP}$) in the such device can be found in Ref. [36] as a function of the gap thickness, together with a thorough analysis of why quenching does not impose severe limitations on the quantum efficiency.

**Conclusion**
We have developed a general method for computing far-field free-space and guided-mode radiation diagrams of local inhomogeneities in stratified media. The far-field retrieval just requires the knowledge of the near-zone electromagnetic field (obtained by a Maxwell's equation solver) on a closed surface that envelopes the local inhomogeneity. Moreover, the retrieval is based on rigorous mode decomposition. The accuracy of the tool has been corroborated by comparisons between fully numerical results and analytical results for two simple examples, and its range of applicability has been further illustrated with two examples taken from the recent literature on nanoplasmonics. Additionally, we have already successfully applied this numerical tool to various problems, including light extraction with optimized nanoparticles [5], broadband absorbers [37], and emission in metal nanogap structures [36] and metallic patch antennas [38]. Thus we are confident that the tool can be widely used for various light emission and scattering problems in photonics and is compatible with virtually any Maxwell's equation solver.

**Technical remarks**. This section briefly gathers some information on the usage of the numerical tool. (1) **Retrieval procedure.** To retrieve free-space or guided-mode radiation diagram with the Matlab-based open-source software, beforehand one has to calculate the radiated or scattered electromagnetic field with any Maxwell's equation solver on a *rectangular* "box" that *fully* encloses the *local inhomogeneity*. Users have to set the refractive index and thickness of each layer (the on-line version handle nonmagnetic materials only). The numerical tool and its user guide are provided in Ref. [25]. (2) **Guided-mode radiation diagram.** Waveguide mode decomposition requires the profiles of the guided modes in the stratified media, which are found by a mode solver included in the open-source software. In addition, users should provide the modal effective index $n_{eff}$ (potentially a c-number) and set the polarization of the desired mode. Accurate value of $n_{eff}$ is not required, since it is computed precisely by the mode solver. Note that the $n_{eff}$ of cylindrical waveguide mode of

3D stratified structure is the same as the one of the 2D stratified structure (see SI). (3) **CPU time**. The open-source code is rather efficient. For instance, based on an ordinary workstation with a processor main frequency of 1.80 GHz, for the example 1 or 2 in this paper, the retrieval of the free-space radiation diagram (including both upper and lower free space) consumes about 100 seconds, and it just costs 2 seconds to compute the radiation diagrams of both the TE- and TM-modes for the example 2. CPU time is directly determined by the number of sample points on the "box" and the number of far-field directions that are computed.

## ASSOCIATED CONTENT
Supporting Information. (1) Unconjugated form Lorentz reciprocity theorem. (2) Expressions of the cylindrical waveguide mode profiles. (3) Mode orthogonality and normalization. (4) Removing the singularity of Hankel functions at $r = 0$. (5) Relation with the field equivalence principle. This material is available free of charge via the Internet at http://pubs.acs.org.

## AUTHOR INFORMATION
Corresponding Author
*E-mail:
jianji.yang.photonique@gmail.com,
jean-paul.hugonin@institutoptique.fr,
philippe.lalanne@institutoptique.fr

Present Address
[#] Department of Electrical Engineering, Stanford University, Stanford, California 94305, United States.

## ACKNOWLEDGMENTS
Part of the study was carried out with financial support from "the Investments for the future" Programme IdEx Bordeaux–LAPHIA (ANR-10-IDEX-03-02)

# Supporting Information for
# "Near-to-far field transformations for radiative and guided waves"

Jianji Yang[1], Jean-Paul Hugonin[2] and Philippe Lalanne[1]

1 Laboratoire Photonique Numérique et Nanosciences (LP2N), Institut d'Optique d'Aquitaine, Université Bordeaux, CNRS, 33400 Talence, France.
2 Laboratoire Charles Fabry, Institut d'Optique Graduate School, CNRS, Univ. Paris-Sud, 2 Avenue Augustin Fresnel, 91127 Palaiseau, France
E-mail:   jianji.yang.photonique@gmail.com
          jean-paul.hugonin@institutoptique.fr
          philippe.lalanne@institutoptique.fr

## Content

**1. Unconjugated form Lorentz reciprocity theorem**
**2. Expressions of the cylindrical waveguide mode profiles**
**3. Mode orthogonality and normalization**
**4. Removing the singularity of Hankel functions at *r* = 0**
**5. Relation with the field equivalence principle**

## 1. Unconjugated form of Lorentz reciprocity theorem

A key basis of the present work is the unconjugated form of Lorentz reciprocity theorem [1], which is repeatedly used in the work and hence derived here. Assuming the stack materials are reciprocal, we consider two solutions (labeled by the subscripts 1 and 2) of Maxwell's equations,

$$\nabla \times \mathbf{E}_1 = i\omega_1\mu_1\mathbf{H}_1 + \mathbf{M}_1 \text{ and } \nabla \times \mathbf{H}_1 = -i\omega_1\varepsilon_1\mathbf{E}_1 + \mathbf{J}_1, \tag{S1a}$$

$$\nabla \times \mathbf{E}_2 = i\omega_2\mu_2\mathbf{H}_2 + \mathbf{M}_2 \text{ and } \nabla \times \mathbf{H}_2 = -i\omega_2\varepsilon_2\mathbf{E}_2 + \mathbf{J}_2. \tag{S1b}$$

By applying the divergence theorem to $\mathbf{E}_2 \times \mathbf{H}_1 - \mathbf{E}_1 \times \mathbf{H}_2$, the reciprocity theorem can be found as

$$\oiint_\Sigma (\mathbf{E}_2 \times \mathbf{H}_1 - \mathbf{E}_1 \times \mathbf{H}_2) \bullet d\mathbf{S}$$
$$= \iiint_\Omega [i\mathbf{H}_1 \cdot \mathbf{H}_2(\mu_2\omega_2 - \mu_1\omega_1) + i\mathbf{E}_1 \cdot \mathbf{E}_2(\varepsilon_1\omega_1 - \varepsilon_2\omega_2) + \mathbf{H}_1 \cdot \mathbf{M}_2 - \mathbf{E}_2 \cdot \mathbf{J}_1 - \mathbf{H}_2 \cdot \mathbf{M}_1 + \mathbf{E}_1 \cdot \mathbf{J}_2]dV, \tag{S2a}$$

where $\Sigma$ denotes a surface enclosing the domain $\Omega$, which contains the source terms ($\mathbf{J}_1$, $\mathbf{J}_2$, $\mathbf{M}_1$, and $\mathbf{M}_2$). If the domain $\Omega$ is source-free, Eq. (S2a) becomes

$$\oiint_\Sigma (\mathbf{E}_2 \times \mathbf{H}_1 - \mathbf{E}_1 \times \mathbf{H}_2) \bullet d\mathbf{S} = 0. \tag{S2b}$$

## 2. Expression of the waveguide mode profiles in cylindrical coordinates

Generally, when solving the source-free Maxwell's equations of a stratified waveguide in a cylindrical coordinate $(r,\varphi,z)$, the mode profile can be casted into independent functions of $z-$, $\varphi-$, and $r-$. The $z$-dependent profiles correspond to the usual mode profiles obtained in Cartesian coordinates that are derived in all textbooks with an invariance along one of the longitudinal directions , see Fig. S1. Due to the revolution symmetry, the $\varphi$-dependence is simply $\exp(in\varphi)$ ($n$ = 0, ±1 …). And finally, the $r$-dependence is characterized by Hankel functions. The propagation constant (along the radial direction $r$) is the same as the one obtained in Cartesian coordinate.

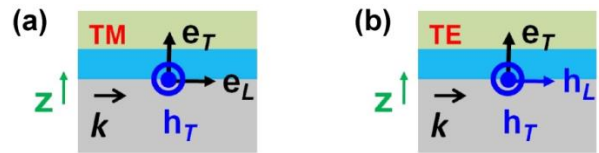

**Figure S1. Field components of the modes of a stratified waveguide in Cartesian coordinates** (all fields components are assumed to be independent of the coordinate along the direction perpendicular to the figure). **(a)** and **(b)** TM and TE modes. For a TM mode, the field component triplet is ($\mathbf{e}_T$, $\mathbf{e}_L$, $\mathbf{h}_T$); and for a TE mode, the triplet is ($\mathbf{e}_T$, $\mathbf{h}_L$, $\mathbf{h}_T$). '$T$' and '$L$' label the transverse and longitudinal components. $\mathbf{k}$ indicates the propagation constant and the mode profiles are only $z$-dependent. All details can be found in Ref. [2] for instance.

Throughout the work, we denote the $n^{th}$ *azimuthal harmonic* of the $m^{th}$ outgoing (labeled by "+") and ingoing (labeled by "−") waveguide modes as $\Phi_{+m,n} = \left(\hat{\mathbf{E}}^+_{m,n}\exp(in\varphi), \hat{\mathbf{H}}^+_{m,n}\exp(in\varphi)\right)$ and $\Phi_{-m,n} = \left(\hat{\mathbf{E}}^-_{m,n}\exp(in\varphi), \hat{\mathbf{H}}^-_{m,n}\exp(in\varphi)\right)$, where $\hat{\mathbf{E}}^{\pm}_{m,n}$ ($\hat{\mathbf{H}}^{\pm}_{m,n}$) depends on $z$ and $r$.

The explicit expressions of TM- and TE-mode profiles and their asymptotic expressions as $r \to \infty$ are provided in the following Eqs. (S3a)-(S3d), where we drop $\exp(\pm in\varphi)$ for the sake of simplicity and organize the field components with the usual sequence $(e_r, e_\varphi, e_z, h_r, h_\varphi, h_z)$.

(a) TM outgoing modes

$$\begin{bmatrix}\hat{\mathbf{E}}^+_{m,n} \\ \hat{\mathbf{H}}^+_{m,n}\end{bmatrix} = \begin{bmatrix} \frac{i}{2}\left(H^+_{n+1}(k_m r) - H^+_{n-1}(k_m r)\right)\mathbf{e}_L(z) \\ \frac{1}{2}\left(H^+_{n+1}(k_m r) + H^+_{n-1}(k_m r)\right)\mathbf{e}_L(z) \\ H^+_n(k_m r)\mathbf{e}_T(z) \\ \frac{-1}{2}\left(H^+_{n+1}(k_m r) + H^+_{n-1}(k_m r)\right)\mathbf{h}_T(z) \\ \frac{i}{2}\left(H^+_{n+1}(k_m r) - H^+_{n-1}(k_m r)\right)\mathbf{h}_T(z) \\ 0 \end{bmatrix} \xrightarrow{r \to \infty} \sqrt{\frac{2}{\pi k_m r}} \exp\left[i\left(k_m r - \frac{2n+1}{4}\pi\right)\right] \begin{bmatrix} \mathbf{e}_L(z) \\ 0 \\ \mathbf{e}_T(z) \\ 0 \\ \mathbf{h}_T(z) \\ 0 \end{bmatrix}, \quad (S3a)$$

(b) TM ingoing modes

$$\begin{bmatrix}\hat{\mathbf{E}}^-_{m,n} \\ \hat{\mathbf{H}}^-_{m,n}\end{bmatrix} = \begin{bmatrix} \frac{i}{2}\left(H^-_{-n+1}(k_m r) - H^-_{-n-1}(k_m r)\right)\mathbf{e}_L(z) \\ \frac{1}{2}\left(H^-_{-n+1}(k_m r) + H^-_{-n-1}(k_m r)\right)\mathbf{e}_L(z) \\ H^-_{-n}(k_m r)\mathbf{e}_T(z) \\ \frac{-1}{2}\left(H^-_{-n+1}(k_m r) + H^-_{-n-1}(k_m r)\right)\mathbf{h}_T(z) \\ \frac{i}{2}\left(H^-_{-n+1}(k_m r) - H^-_{-n-1}(k_m r)\right)\mathbf{h}_T(z) \\ 0 \end{bmatrix} \xrightarrow{r \to \infty} \sqrt{\frac{2}{\pi k_m r}} \exp\left[-i\left(k_m r + \frac{2n-1}{4}\pi\right)\right] \begin{bmatrix} -\mathbf{e}_L(z) \\ 0 \\ \mathbf{e}_T(z) \\ 0 \\ -\mathbf{h}_T(z) \\ 0 \end{bmatrix}, \quad (S3b)$$

(c) TE outgoing modes

$$\begin{bmatrix}\hat{\mathbf{E}}^+_{m,n} \\ \hat{\mathbf{H}}^+_{m,n}\end{bmatrix} = \begin{bmatrix} \frac{-1}{2}\left(H^+_{n+1}(k_m r) + H^+_{n-1}(k_m r)\right)\mathbf{e}_T(z) \\ \frac{i}{2}\left(H^+_{n+1}(k_m r) - H^+_{n-1}(k_m r)\right)\mathbf{e}_T(z) \\ 0 \\ \frac{i}{2}\left(H^+_{n+1}(k_m r) - H^+_{n-1}(k_m r)\right)\mathbf{h}_L(z) \\ \frac{1}{2}\left(H^+_{n+1}(k_m r) + H^+_{n-1}(k_m r)\right)\mathbf{h}_L(z) \\ H^+_n(k_m r)\mathbf{h}_T(z) \end{bmatrix} \xrightarrow{r \to \infty} \sqrt{\frac{2}{\pi k_m r}} \exp\left[i\left(k_m r - \frac{2n+1}{4}\pi\right)\right] \begin{bmatrix} 0 \\ \mathbf{e}_T(z) \\ 0 \\ \mathbf{h}_L(z) \\ 0 \\ \mathbf{h}_T(z) \end{bmatrix}, \quad (S3c)$$

(d) TE ingoing modes

$$\begin{bmatrix} \hat{\mathbf{E}}^-_{m,n} \\ \hat{\mathbf{H}}^-_{m,n} \end{bmatrix} = \begin{bmatrix} \frac{-1}{2}\left(H^-_{-n+1}(k_m r)+H^-_{-n-1}(k_m r)\right)\mathbf{e}_T(z) \\ \frac{i}{2}\left(H^-_{-n+1}(k_m r)-H^-_{-n-1}(k_m r)\right)\mathbf{e}_T(z) \\ 0 \\ \frac{i}{2}\left(H^-_{-n+1}(k_m r)-H^-_{-n-1}(k_m r)\right)\mathbf{h}_L(z) \\ \frac{1}{2}\left(H^-_{-n+1}(k_m r)+H^-_{-n-1}(k_m r)\right)\mathbf{h}_L(z) \\ H^-_{-n}(k_m r)\mathbf{h}_T(z) \end{bmatrix} \xrightarrow{r\to\infty} \sqrt{\frac{2}{\pi k_m r}}\exp\left[-i\left(k_m r+\frac{2n-1}{4}\pi\right)\right]\begin{bmatrix} 0 \\ -\mathbf{e}_T(z) \\ 0 \\ -\mathbf{h}_L(z) \\ 0 \\ \mathbf{h}_T(z) \end{bmatrix}. \quad \text{(S3d)}$$

In Eqs. (S3a)-(S3d), $\mathbf{e}_L(z)$ and $\mathbf{e}_T(z)$ ($\mathbf{h}_L(z)$ and $\mathbf{h}_T(z)$) represent the longitudinal (labeled by "*L*") and transverse (labeled by "*T*") electric (magnetic) components of the *z*-dependent mode profiles, which correspond to the classical mode profiles calculated in Cartesian coordinates, see Fig. S1; $H^+_n(k_m r)$ and $H^-_n(k_m r)$ denote the Hankel function of the first and second kind of order *n*. The asymptotic expressions are derived using the asymptotic forms of Hankel functions [3].

### 3. Unconjugated form of mode orthogonality and normalization

The mode decomposition technique formulated in the main text relies on mode orthogonality and normalization relations that are established here.

We consider two cylindrical modes, the $n^{th}$ azimuthal harmonic of the $m^{th}$ mode $\Phi_{\sigma m,n} = \left(\hat{\mathbf{E}}^\sigma_{m,n}, \hat{\mathbf{H}}^\sigma_{m,n}\right)\exp(\sigma i n\varphi)$ and the $q^{th}$ azimuthal harmonic of the $p^{th}$ mode $\Phi_{\xi p,q} = \left(\hat{\mathbf{E}}^\xi_{p,q}, \hat{\mathbf{H}}^\xi_{p,q}\right)\exp(\xi i q\varphi)$, with $\sigma = \pm 1$ and $\xi = \pm 1$ dictating the mode propagation direction (with '+' and '–' corresponding to outgoing and ingoing modes, respectively).

We define the *inner product* of two modes as

$$\Phi_{\sigma m,n}\otimes\Phi_{\xi p,q} = \int_0^{2\pi}\int_{-\infty}^{+\infty}\exp\left[i\varphi(\sigma n+\xi q)\right]\left(\hat{\mathbf{E}}^\sigma_{m,n}\times\hat{\mathbf{H}}^\xi_{p,q}-\hat{\mathbf{E}}^\xi_{p,q}\times\hat{\mathbf{H}}^\sigma_{m,n}\right)\bullet\mathbf{n}_r r\,dz\,d\varphi,\ (\sigma=\pm 1,\xi=\pm 1) \quad \text{(S4)}$$

where $\otimes$ denotes the *inner product operator* and $\mathbf{n}_r$ is the normalized radial vector of the cylindrical basis. Note that the integral in Eq. (S4) runs over a cylindrical surface of radius *r* from $z=-\infty$ to $z=+\infty$, such as the surfaces $\Sigma_1$ or $\Sigma_2$ sketched in Fig. S2. If two modes have different polarizations, they are orthogonal to each other because the cross-products are null. In **Section 3.2**, we additionally show the *orthogonality relation* for two modes with the same polarization

$$\Phi_{\sigma m,n}\otimes\Phi_{\xi p,q} = \delta_{\sigma m,-\xi p}\delta_{n,q}N_{m,n}, \quad \text{(S5)}$$

where $N_{m,n}$ denotes the *normalization coefficient* of mode $\Phi_{+m,n}$ (see **Section 3.3**)

$$N_{m,n} = \Phi_{+m,n}\otimes\Phi_{-m,n} = (-1)^n\frac{8}{k_m}\int_{-\infty}^{+\infty}\left(\mathbf{e}_{m,T}\times\mathbf{h}_{m,T}\right)\bullet\mathbf{n}_r\,dz. \quad \text{(S6)}$$

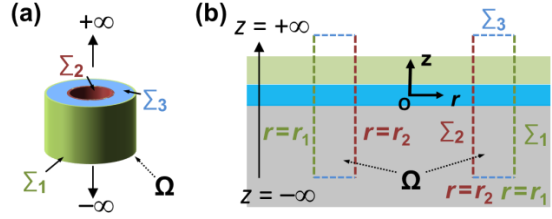

**Figure S2. Sketch of a cylindrical-shell domain Ω. (a)** Domain Ω and its outer surface $\Sigma_1$ (green), inner surface $\Sigma_2$ (red); the surface $\Sigma_3$ (blue) is situated at $|z| = +\infty$.

**(b)** Cross-sectional view. Cylindrical surfaces $\Sigma_1$ ($r = r_1$, vertical green lines) and $\Sigma_2$ ($r = r_2$, vertical red lines) are the outer and inner surfaces of the compact domain Ω. $\Sigma_3$ (shown with four horizontal blue lines) presents the horizontal boundaries of Ω at $|z| = +\infty$.

### 3.1 Radial independence of Eqs. (S5) and (S6) for guided modes

In this sub-section, we will demonstrate the *r*-independence property of Eqs. (S5) and (S6); this independence is important for establishing the mode orthogonality and normalization in **Section 3.2**.

Yet we consider two modes, $\Phi_{\sigma m,n}$ and $\Phi_{\xi p,q}$ ($\sigma = \pm 1$ and $\xi = \pm 1$, labeling the propagating direction), and a cylindrical-shell volume Ω shown in Fig. S2. Since modes are solution of source-free Maxwell's equations, we can inject $\Phi_{\sigma m,n}$ [as solution 1 ($\mathbf{E}_1, \mathbf{H}_1$)] and $\Phi_{\xi p,q}$ [as solution 2 ($\mathbf{E}_2, \mathbf{H}_2$)] into Eq. (S2b) (Lorentz reciprocity theorem) for the source-free domain Ω. We easily obtain $\iint_{\Sigma_1+\Sigma_2+\Sigma_3} (\mathbf{E}_2 \times \mathbf{H}_1 - \mathbf{E}_1 \times \mathbf{H}_2) \bullet d\mathbf{S} = 0$. Furthermore, if the modes are truly guided modes that are exponentially damped at $|z| \to \infty$, the integral over $\Sigma_3$ is null and hence we obtain

$$\iint_{\Sigma_1} (\mathbf{E}_2 \times \mathbf{H}_1 - \mathbf{E}_1 \times \mathbf{H}_2) \bullet d\mathbf{S} = -\iint_{\Sigma_2} (\mathbf{E}_2 \times \mathbf{H}_1 - \mathbf{E}_1 \times \mathbf{H}_2) \bullet d\mathbf{S}, \quad (S7)$$

the minus sign resulting from the orientation of the normal vectors of the surfaces. According to Eq. (S4), it is straightforward to find that

$$\Phi_{\sigma m,n} \otimes \Phi_{\xi p,q}\big|_{r_1} = \iint_{\Sigma_1} (\mathbf{E}_2 \times \mathbf{H}_1 - \mathbf{E}_1 \times \mathbf{H}_2) \bullet d\mathbf{S}, \ (\sigma = \pm 1 \text{ and } \xi = \pm 1) \quad (S8a)$$

$$\Phi_{\sigma m,n} \otimes \Phi_{\xi p,q}\big|_{r_2} = -\iint_{\Sigma_2} (\mathbf{E}_2 \times \mathbf{H}_1 - \mathbf{E}_1 \times \mathbf{H}_2) \bullet d\mathbf{S}, \quad (S8b)$$

where $\Phi_{\sigma m,n} \otimes \Phi_{\xi p,q}\big|_{r_1}$ and $\Phi_{\sigma m,n} \otimes \Phi_{\xi p,q}\big|_{r_2}$ represent inner products performed at cylindrical surfaces $r = r_1$ and $r = r_2$. By substituting Eqs. (S8a) and (S8b) into Eq. (S7), we find that

$$\Phi_{\theta m,n} \otimes \Phi_{\xi p,q}\big|_{r_1} = \Phi_{\sigma m,n} \otimes \Phi_{\xi p,q}\big|_{r_2}, \tag{S9}$$

indicating that the orthogonality and normalization relations given by Eqs. (S5) and (S6) are *r*-independent.

## 3.2. Mode orthogonality relation

As $\Phi_{\sigma m,n} \otimes \Phi_{\xi p,q}$ ($\sigma = \pm 1$ and $\xi = \pm 1$, labeling the propagating direction) is *r*-independent, for simplicity we may derive the mode orthogonality relation Eq. (S5) by considering $r \to \infty$. Referring to Eqs. (S3a)-(S3d), we calculate $\Phi_{\sigma m,n} \otimes \Phi_{\xi p,q}$ for $r \to \infty$ using the asymptotic expressions for the modes

$$\Phi_{\sigma m,n} \otimes \Phi_{\xi p,q} = \frac{2}{\pi\sqrt{k_m k_p}} \exp[i(\sigma k_m + \xi k_p)r]\exp[-i\pi(2(n+q)+\sigma+\xi)/4] \times$$
$$\int_0^{2\pi} \exp[i\varphi(\sigma n + \xi q)]d\varphi \int_{-\infty}^{+\infty} (\xi \mathbf{e}_{m,T} \times \mathbf{h}_{p,T} - \sigma \mathbf{e}_{p,T} \times \mathbf{h}_{m,T}) \bullet \mathbf{n}_r dz \tag{S10}$$

To show that Eq. (S10) leads to Eq. (S5), we consider several cases:

(**a**) Different mode polarization: From the asymptotic expressions of TM and TE modes, one easily finds $\hat{\mathbf{E}}_{m,n}^{TE} \times \hat{\mathbf{H}}_{p,q}^{TM} - \hat{\mathbf{E}}_{p,q}^{TM} \times \hat{\mathbf{H}}_{m,n}^{TE} = 0$. Therefore, $\Phi_{\sigma m,n}^{TE} \otimes \Phi_{\xi p,q}^{TM} = 0$ for two modes with different polarizations.

(**b**) Different azimuthal indices: For $|n| \neq |q|$, $\int_0^{2\pi} \exp[i\varphi(\sigma n + \xi q)]d\varphi = 0$ and hence $\Phi_{\sigma m,n} \otimes \Phi_{\xi p,q} = 0$.

(**c**) Different propagation constant ($k_m \neq k_p$): Note that in Eqs. (S3a)-(S3d), the $m^{th}$ outgoing ($\sigma = +1$) and ingoing ($\sigma = -1$) modes have the same value of propagation constant $k_m$, but their propagation directions (i.e., being outgoing or ingoing) is discriminated by the notation '±' in front of $k_m$. When $k_m \neq k_p$, a simple relation $\int_{-\infty}^{+\infty} (\xi \mathbf{e}_{m,T} \times \mathbf{h}_{p,T} - \sigma \mathbf{e}_{p,T} \times \mathbf{h}_{m,T}) \bullet \mathbf{n}_r dz = 0$ can be found [4], which is the unconjugated form of mode orthogonality relation for planar waveguides in a Cartesian coordinate, and it ensures $\Phi_{\sigma m,n} \otimes \Phi_{\xi p,q} = 0$.

(**d**) The same value of propagation constant $k_m = k_p$ and the same propagation direction (i.e., $\sigma = \xi$): In this case, we have

$$\int_{-\infty}^{+\infty} (\xi \mathbf{e}_{m,T} \times \mathbf{h}_{p,T} - \sigma \mathbf{e}_{p,T} \times \mathbf{h}_{m,T}) \bullet \mathbf{n}_r dz = \sigma \int_{-\infty}^{+\infty} (\mathbf{e}_{p,T} \times \mathbf{h}_{p,T} - \mathbf{e}_{p,T} \times \mathbf{h}_{p,T}) \bullet \mathbf{n}_r dz = 0,$$

and it again ensures $\Phi_{\sigma m,n} \otimes \Phi_{\xi p,q} = 0$.

## 3.3. Mode normalization

To normalize an outgoing mode $\Phi_{+m,n} = (\hat{\mathbf{E}}^+_{m,n}, \hat{\mathbf{H}}^+_{m,n})\exp(in\varphi)$, we use the counter-propagative mode $\Phi_{-m,n} = (\hat{\mathbf{E}}^-_{m,n}, \hat{\mathbf{H}}^-_{m,n})\exp(-in\varphi)$. By substituting the two modes into Eq. (S5), we find that

$$\Phi_{+m,n} \otimes \Phi_{-m,n} = (-1)^{n+1} \frac{8}{k_m} \int_{-\infty}^{+\infty} (\mathbf{e}_{m,T} \times \mathbf{h}_{m,T}) \bullet \mathbf{n}_r dz. \tag{S11}$$

It is convenient to normalize the modes such that

$$\int_{-\infty}^{+\infty} (\mathbf{e}_{m,T} \times \mathbf{h}_{m,T}) \bullet \mathbf{n}_r dz = -2, \tag{S12}$$

If the planar waveguide is non-lossy [i.e., Im($k_m$) = 0], a planar mode normalized according to Eq. (S12) carries a unitary Poynting flux, i.e. $\frac{1}{2}\text{Re}\left[\int_{-\infty}^{+\infty} (\mathbf{e}_{m,T} \times \mathbf{h}^*_{m,T}) \bullet \mathbf{n}_r dz\right] = 1$.

Injecting Eq. (S12) into Eq. (S11), $\Phi_{m,n}$ is normalized as

$$N_{m,n} = \Phi_{+m,n} \otimes \Phi_{-m,n} = 16(-1)^n / k_m. \tag{S13}$$

### 3.4 Remark on the orthogonality and normalization of radiation modes

In the former sub-sections, mode orthogonality and mode normalization have been established for guided modes, which are of main interest in practice. Here, we make a short remark on the extension of orthogonality and normalization issue to radiation modes. According to Eq. (3) in the main text, the total field is expanded into the waveguide mode basis that includes both a discrete set of truly guided modes (bound states) and a continuum of radiation modes. Therefore, mode orthogonality has to be established for all the modes, especially the orthogonality between bound and continuum states should be established. Such orthogonalities can be derived directly from Maxwell's equations with a formalism based on the Lorentz reciprocity theorem. The formalism and its difficulties to handle leakage of continuum states are documented in textbooks of optical waveguides [4]. Moreover it was generalized to periodic waveguides in [5], by introducing complex spatial coordinate transforms [6] — a sort of perfectly matched layer (PML) — that map the open problem with its associated continuum of radiation states to an approximated closed problem with a countable number of discrete states, called quasi-normal Bloch modes. For brevity, the introduction of complex spatial coordinate transforms allows one to handle the orthogonality and normalization issues for both guided and radiation modes of stratified waveguides that is treated in the current work. The PML-assisted approach has been effectively applied in several theoretical studies on waveguides, such as the computation of scattering loss at waveguide terminations [7], the coupling of quantum emitters with photonic crystal waveguides [8], and the localization lengths of periodic waveguides with tiny imperfections [9].

### 4. Removing the singularity of Hankel functions at $r = 0$

Hankel function $H_n^\pm(r) = J_n(r) \pm iY_n(r)$ is a linear combination of Bessel functions of the first $J_n(r)$ and second $Y_n(r)$ kinds [3]. $J_n(r)$ is finite at $r = 0$, however $Y_n(r)$ diverges at $r = 0$ and therefore the cylindrical modes (including both the outgoing mode $\Phi_{+m,n}$ and ingoing mode $\Phi_{-m,n}$) expressed by Hankel functions have singularity at $r = 0$. This results in difficulty to numerically implement the calculations of Eqs. (8)-(11) in the main text. Note that, $H_n^+(r) + H_n^-(r) = 2J_n(r)$ is finite and therefore $\Phi_{+m,n} + \Phi_{-m,n}$ does not diverge at $r = 0$. According to the mode orthogonality relation given by Eq. (7) in the main text or Eq. (S5), one can easily find $\Phi_{+m,n} \otimes \Phi_{+m,n} = 0$ and then

$$N_{m,n} = \Phi_{+m,n} \otimes \Phi_{-m,n} = \Phi_{+m,n} \otimes (\Phi_{-m,n} + \Phi_{+m,n}). \tag{S14}$$

So in Eq. (8) of the main text, $\Phi_{-m,n}$ can be replaced by $\Phi_{+m,n} + \Phi_{-m,n}$ to compute the excitation coefficient, *i.e.*, $c_{m,n}^+ = (\mathbf{E},\mathbf{H}) \otimes \Phi_{-m,n} = (\mathbf{E},\mathbf{H}) \otimes (\Phi_{+m,n} + \Phi_{-m,n})$. Furthermore, this important replacement is applied into Eqs. (8)-(11) of the main text to avoid any singularity in numerical calculations.

## 5. Relation with the field equivalence principle

The field equivalence principle classically used for NFFT [1] has not been used explicitly in the present derivations. Nevertheless, it is helpful for understanding the proposed methods, e.g., Eqs. (2) and (11) in the main text, to establish the link between the present method and the field equivalent principle. Let's take Eq. (11) as an example, with some simple vector identities, one can rewrite Eq. (11) as

$$c_{m,n}^+ = \frac{-1}{N_{m,n}} \oint_{\Sigma_B} \exp(-in\varphi)(\hat{\mathbf{E}}_{m,n}^- \bullet \mathbf{J} - \hat{\mathbf{H}}_{m,n}^- \bullet \mathbf{M}) dS, \tag{S15}$$

with electric *surface* current $\mathbf{J} = -\mathbf{n} \times \mathbf{H}$ and magnetic *surface* current $\mathbf{M} = \mathbf{n} \times \mathbf{E}$. Converting Eq. (11) to Eq. (S15) amounts to casting $(\mathbf{E},\mathbf{H})$ on $\Sigma_B$ into fictitious sources according to the field equivalent principle. Note that the amplitude of an outgoing mode excited by local sources is determined by the corresponding ingoing mode and the same sources, as the modes conforms to the unconjugated form of orthogonality, see for instance the dipole emission in a periodic waveguide in Ref. [5].